\RequirePackage{fix-cm}
\documentclass[conference]{IEEEtran}

\usepackage{graphicx}
\usepackage{color}

\usepackage[numbers]{natbib}

\usepackage{hyperref}

\newcommand{\e}[1]{\mbox{\lstinline[basicstyle=\normalsize]|#1|}}

\usepackage{listings}
\usepackage{eiffel}
\usepackage{boogie}

\usepackage{amsmath,amssymb}
\usepackage{wasysym}
\usepackage{enumerate}
\usepackage{color}      
\usepackage{amssymb}    

\begin{document}

\title{A contract-based method to specify stimulus-response requirements}



\author{\IEEEauthorblockN{Alexandr Naumchev,
Manuel Mazzara, Bertrand Meyer}
\IEEEauthorblockA{Innopolis University\\
Innopolis, Russian Federation\\
\{a.naumchev, m.mazzara, b.meyer\}@innopolis.ru\\
Jean-Michel Bruel, Florian Galinier, Sophie Ebersold\\
Toulouse University\\
Toulouse, France\\
\{bruel, galinier, ebersold\}@irit.fr}}




\maketitle

\begin{abstract}
A number of formal methods exist for capturing stimulus-response requirements in a declarative form. Someone yet needs to translate the resulting declarative statements into imperative programs. The present article describes a method for specification and verification of stimulus-response requirements in the form of imperative program routines with conditionals and assertions. A program prover then checks a candidate program directly against the stated requirements. The article illustrates the approach by applying it to an ASM model of the Landing Gear System, a widely used realistic example proposed for evaluating specification and verification techniques.
\end{abstract}

\begin{IEEEkeywords}
Seamless Requirements, Design by Contract, AutoProof, Eiffel, Landing Gear System
\end{IEEEkeywords}

\section{Overview and Main Results} \label{sec:overview}

The present article describes a technique for specification and verification of stimulus-response requirements using a general-purpose programming language (Eiffel) and a program prover (AutoProof \cite{tschannen2015autoproof}) based on the principles of Design by Contract \cite{meyer2009touch}. 

Real-time, or reactive, systems are often run by a software controller that repeatedly executes one and the same routine and it is specified to take actions at specific time intervals or according to external stimuli \cite{Hayes2003}. This architecture is reasonable when the software has to react timely to non-deterministic changes in the environment. In this case the program should react to the external stimuli in small steps, so that in the event of a new change it responds timely.

Computation tree logics (CTL) \cite{clarke1982design} represent a frequent choice when it comes to capturing stimulus-response requirements. Although it may be easier to reason about requirements using declarative logic like CTL, the reasoning may be of little value for the software developer who will implement the requirements.  
Mainstream programming languages are all imperative, and the translation between declarative requirements and imperative programs is semi-formal. 

Requirements have to be of imperative 
nature from the beginning. This would bridge the gap in how customers and developers understand them. For a software developer it is preferable to reason about the future program without switching to an additional formalism, notation and tools not connected to the original programming language and the IDE. The present article describes a technique to achieve this goal, in particular:

\begin{itemize}
\item Introduces the Landing Gear System (LGS) case study and the LGS baseline requirements (\autoref{sec:landing_gear_example}).
\item Generalizes the LGS baseline requirements, maps them to a well-established taxonomy, and complements the taxonomy (\autoref{sec:stim_resp_reqs}).
\item Provides a general scheme for capturing semantics of the stimulus-response requirements in the form of imperative program routines with assertions (\autoref{sec:reqs_translation}).
\item Exercises utility of the approach by applying it to an Abstract State Machine (ASM) specification of the Landing Gear System case study (\autoref{sec:applying_to_example}).
\item Concludes the possibility of statically checking a sequential imperative program directly against a stimulus-response requirement whose semantics is expressed in the same programming language through conditionals, loops, and assertions (\autoref{sec:conclusions}).
\end{itemize}

Application of the technique leads to discovery of an error in the published model of the LGS ASM \cite{arcaini2014modeling}. The error is not present in the specification the authors have actually used for proving the properties, but the error has found its way into the publication.

\section{The Landing Gear System}
\label{sec:landing_gear_example}

\begin{figure}
\includegraphics[scale=0.64]{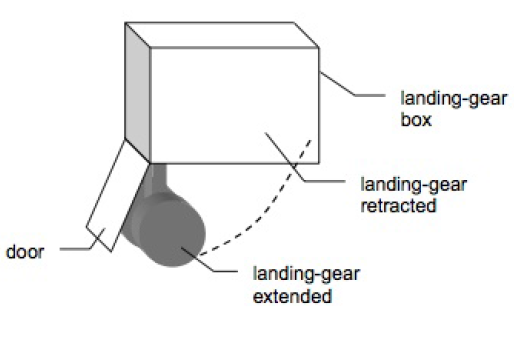}
\caption{Landing set (source: \cite{boniol2014landing}).}
\label{fig:lgs}
\end{figure}

Landing Gear System was proposed as a benchmark for techniques and tools dedicated to the verification of behavioral properties of systems \cite{boniol2014landing}. It physically consists of the landing set, a gear box that stores the gear in the retracted position, and a door attached to the box (\autoref{fig:lgs}). The door and the gear are actuated independently by a digital controller. The controller reacts to changes in position of a handle in the cockpit by initiating either gear extension or retraction process. The task is to program the controller so that it correctly aligns in time the events of changing the handle's position and sending commands to the door and the gear actuators.

\begin{figure}
\begin{description}
\item[$(R_{11}bis)$] When the command line is working (normal mode), if the landing gear command handle has been pushed DOWN and stays DOWN, then eventually the gears will be locked down and the doors will be seen closed.
\item[$(R_{12}bis)$] When the command line is working (normal mode), if the landing gear command handle has been pushed UP and stays UP, then eventually the gears will be locked retracted and the doors will be seen closed.
\item[$(R_{21})$] When the command line is working (normal mode), if the landing gear command handle remains in the DOWN position, then retraction sequence is not observed.
\item[$(R_{22})$] When the command line is working (normal mode), if the landing gear command handle remains in the UP position, then outgoing sequence is not observed.
\end{description}
\caption{Baseline LGS requirements.}
\label{fig:baseline_requirements}
\end{figure}

\section{Stimulus-Response requirements} \label{sec:stim_resp_reqs}
The LGS case study defines a number of requirements, including several for the normal mode of operation (\autoref{fig:baseline_requirements}).
The requirements communicate a common meaning of the form: 
\begin{itemize}
\item If \textit{stimulus} holds, then \textit{response} will eventually hold in the future.
\end{itemize}
For requirement $R_{11}bis$,
\begin{equation*}
\begin{aligned}
stimulus \Leftrightarrow &``The\ operation\ mode\ is\ normal\ and\\
&the\ handle\ is\ DOWN"
\end{aligned}
\end{equation*}
and
\begin{equation*}
\begin{aligned}
response \Leftrightarrow (stimulus \implies &``The\ gears\ are\ down\ and\\
&the\ doors\ are\ closed")
\end{aligned}
\end{equation*}
The implication in the definition of $response$ reflects the ``and stays DOWN" part of the original requirement.

In addition to that, requirements $R_{21}$ and $R_{22}$ communicate something else:
\begin{itemize}

\item Once \textit{response} holds in the presence of \textit{stimulus}, and \textit{stimulus} holds forever, \textit{response} will hold forever.
\end{itemize}

\subsection{Temporal interpretation of the requirements} \label{sec:sec:temporal_requirements}

The authors of the LGS ASM specification start with a ground model that satisfies a subset of requirements, and then refine the model to satisfy more requirements. The present article focuses on their ground model and the corresponding baseline requirements it covers (\autoref{fig:baseline_requirements}). The work expresses the baseline requirements as CTL properties. The CTL interpretation assigns precise meanings to the requirements by assuming small-step execution semantics of ASM's. In particular, for requirements $R_{11}bis$ and $R_{12}bis$ ``the future'' means ``after a finite number of execution steps'', while for $R_{21}$ and $R_{22}$ ``the future'' means ``after one execution step''.

The finite number of steps in $R_{11}bis$ and $R_{12}bis$ may be unacceptably large though for a system like an LGS of an aircraft. In particular, flights have some expected durations, and the gears have to react to commands in some limited time frame as well. The following two major categories of stimulus-response requirements stem from the speculations above:
\begin{itemize}
\item \textit{If stimulus holds, then response will hold in not more than k execution steps.}\\
Requirements of this form are also called \textbf{maximal distance} requirements \cite{koymans1990specifying}.
\item \textit{If stimulus holds, then response will hold in exactly k execution steps.}\\
Requirements of this form are also called \textbf{exact distance}, or \textit{delay} requirements.
\end{itemize}

These two categories are not enough though for capturing stimulus-response requirements. For example, if according to $R_{11}bis$ the gears are locked down and the doors seen closed as the result of the handle staying down, we want this state to be stable if the handle stays down. This leads us to stimulus-response requirements of the following form:
\begin{itemize}
\item \textit{If response holds under stimulus, it will still hold after one execution step in the presence of that stimulus.}\\
Let us call such requirements \textbf{response stability} requirements.
\end{itemize}

\begin{figure}
\begin{description}
\item[$(R_{11}rs)$] If the gears are locked extended and the doors are closed when the landing gear command handle is DOWN, this state will still hold if the handle stays DOWN.
\item[$(R_{12}rs)$] If the gears are locked retracted and the doors are closed when the landing gear command handle is UP, this state will still hold if the handle stays UP.
\end{description}
\caption{LGS response stability requirements.}
\label{fig:rs_requirements}
\end{figure}

It makes sense to complement requirements $(R_{11}bis)$ and $(R_{12}bis)$ with the corresponding response stability requirements (\autoref{fig:rs_requirements}): not only do we want the LGS to respond to a change in the handle's position, but we also want it to maintain the response if the position does not change.

\section{Translation of Stimulus-Response Requirements} \label{sec:reqs_translation}

Assuming the presence of an infinite loop \e{from until False loop main end} that runs a reactive system, a temporal stimulus-response requirement (\autoref{sec:sec:temporal_requirements}) takes the form of a routine with an assertion (\e{check end} construct in Eiffel). The authors draw this idea from the notion of a specification driver \cite{anaumchev2016drivers} - a contracted routine that forms a proof obligation in Hoare logic. AutoProof is a prover
of Eiffel programs that makes it possible to statically check the assertions.
\subsection{Maximal distance} \label{sec:sec:maximal_distance}

\begin{figure}
\begin{lstlisting}
response_holds_within_k_steps
-- If stimulus holds,
-- response will hold within k steps.
  local
    steps: NATURAL
  do
    if (stimulus) then
      from
        steps := 0
      until
        response or (steps = k)
      loop
        main
        steps := steps + 1
      end
      check
        response
      end
    end
  end
\end{lstlisting}
\caption{Representation of a maximal distance requirement. Regardless of the actual reason for the loop to terminate, the response has to hold if the stimulus held at the entry to the loop.}
\label{fig:response_maximal_distance}
\end{figure}

In the representation of a maximal distance requirement (\autoref{fig:response_maximal_distance}) the ``\e{if stimulus then}" clause captures the presence of the stimulus before the up-to-$k$-length execution fragment, and the ``\e{check response end}'' assertion expresses the need for the response upon completion of the sub-execution. The sub-execution may complete for two possible reasons: either occurrence of the response or consumption of all of the available $k$ steps. In the both cases the response has to hold.

\subsection{Exact distance} \label{sec:sec:exact_distance}

\begin{figure}
\begin{lstlisting}
response_holds_in_k_steps
-- If stimulus holds,
-- response will hold in k steps.
  local
    steps: NATURAL
  do
    if (stimulus) then
      from
        steps := 0
      until
        response or (steps = k)
      loop
        main
        steps := steps + 1
      end
      check
        (response and (steps = k))
      end
    end
  end
\end{lstlisting}
\caption{Representation of an exact distance requirement. Both of the loop exit conditions have to hold for the first time simultaneously if the stimulus held at the entry to the loop.}
\label{fig:response_exact_distance}
\end{figure}

Representation of an exact distance requirement (\autoref{fig:response_exact_distance}) is very similar to that one of a maximal distance, with the ``\e{check (response and (steps = k)) end}'' assertion that makes the difference. Regardless of whether the loop terminates because of \e{response} or \e{steps=k}, the both have to hold upon the termination.

\subsection{Response stability} \label{sec:sec:response_stability}

\begin{figure}
\begin{lstlisting}
response_is_stable_under_stimulus
-- response keeps holding under stimulus.
  do
    if (stimulus and response) then
      main
      check
        (stimulus implies response)
      end
    end
  end
\end{lstlisting}
\caption{Representation of a response stability requirement. If response holds under stimulus in some state, the response should hold in the next state in the presence of the same stimulus.}
\label{fig:stimulus_maintains_state}
\end{figure}

Representation of a response stability requirement (\autoref{fig:stimulus_maintains_state}) says: whenever response holds under stimulus in a state, it will still hold in the presence of the same stimulus in the next state.


\section{Applying the Translation Scheme to the Landing Gear Example}
\label{sec:applying_to_example}

The article exercises the approach on the LGS ASM specification, which is operational by the definition and thus is a subject for translation into an imperative program. For this reason the present section starts with explanation of the rules according to which the authors converted the original specification into an Eiffel program.

\subsection{Translation of ASM specifications} \label{sec:sec:asm_translation}

An ASM specification is a collection of rules taking one of the following three forms \cite{gurevich2000sequential}: assignment (\autoref{sec:sec:sec:assignment}), do-in-parallel (\autoref{sec:sec:sec:do_in_parallel}), and conditional (\autoref{sec:sec:sec:conditional}). If we have general rules for translating these operators into Eiffel then we will be able to translate an arbitrary ASM into an Eiffel program.

\subsubsection{Assignment} \label{sec:sec:sec:assignment}

An ASM assignment looks as follows:
\begin{equation}
f(t_1,..,t_j):=t_0
\label{eq:asm_assignment}
\end{equation}
The semantics is: update the current content of location  $\lambda = (f,(a_1,..,a_j))$, where $a_i$ are values referenced by $t_i$, with the value referenced by $t_0$. 

In Eiffel locations are represented with class attributes, so an ASM's location update corresponds in Eiffel to an attribute assignment.

\subsubsection{Do-in-parallel} \label{sec:sec:sec:do_in_parallel}

An ASM can apply several rules simultaneously in one step:
\begin{equation}
R_1 || ... || R_k
\label{eq:asm_do_in_parallel}
\end{equation}

In order to emulate a parallel assignment in a synchronous setting, one needs to assign first to fresh variables and then assign their values to the original ones. For example, an ASM do-in-parallel statement
\begin{equation}
a,b:=max(a-b, b), min(a-b,b)
\end{equation}
in Eiffel would look like
\begin{lstlisting}
local
  a_intermediate, b_intermediate: INTEGER
do
  a_intermediate := max (a-b, b)
  b_intermediate := min (a-b, b)
  a := a_intermediate
  b := b_intermediate
end
\end{lstlisting}
An attempt to update in parallel identical locations in an ASM corresponds semantically to a crash. The translation scheme not only preserves but strengthens this semantics: an Eiffel program with two local variables declared with identical names will not compile.

\subsubsection{Conditional} \label{sec:sec:sec:conditional}

An ASM conditional \e{if} $t$ \e{then} $R_1$ \e{else} $R_2$ carries the same meaning as in Eiffel, so the translation is straightforward.

\subsection{Ground model} \label{sec:sec:model_trans}

Translation of the original LGS ASM specification into Eiffel is publicly available in a GitHub repository \cite{anaumchev2017lgs} and needs clarification too.

The baseline LGS requirements (\autoref{fig:baseline_requirements}) talk about normal mode of operation. The ground ASM specification captures the normal mode through a model invariant, while the Eiffel translation introduces a special boolean query \e{is_normal_mode} for this purpose. The reason for that is rather technical and has to do with the current limitations in the underlying verification technology. The translation also contains a number of annotations for disabling the complications of the underlying verification methodology \cite{polikarpova2014flexible}. Special comments highlight the annotations and tell explicitly that they have nothing to do with the problem at hand.

The repository contains two versions of the ground model, \e{GROUND_MODEL_ORIGINAL} and \e{GROUND_MODEL}. The original one keeps the error from the ASM model, which is not handling opening doors case in the extension sequence. The second version contains the translation without the error.

\subsection{Requirements}

The two classes include the translations of the baseline requirements plus the response stability requirements introduced in the present article. We do not discuss all of them here: requirements $(R_{11}bis)$ and $(R_{12}bis)$, $(R_{21})$ and $(R_{22})$, $(R_{11}rs)$ and $(R_{12}rs)$ are pairwise similar, which is why we prefer to pick one from each pair.

\begin{figure*}
\begin{lstlisting}
r11_bis
-- If (is_normal_mode and (handle_status = is_handle_down)) hold and remain,
-- ((gear_status = is_gear_extended) and (door_status = is_door_closed)) will hold within 10 steps.
  local
    steps: NATURAL
  do
    if (is_normal_mode and (handle_status = is_handle_down)) then
      from
        steps := 0
      until
        (not (is_normal_mode and (handle_status = is_handle_down))) or
        ((gear_status = is_gear_extended) and (door_status = is_door_closed)) or
        (steps=10)
      loop
        main
        steps := steps + 1
      end
      check
        (not (is_normal_mode and (handle_status = is_handle_down))) or
        ((gear_status = is_gear_extended) and (door_status = is_door_closed))
      end
    end
  end
\end{lstlisting}
\caption{Translation of the ``r11\_bis" requirement.}
\label{fig:r11_bis_translation}
\end{figure*}

\begin{figure*}
\begin{lstlisting}
r21
-- If (is_normal_mode and (handle_status = is_handle_up)) holds and remains,
-- (gear_status /= is_gear_extending) will hold within 1 step.
  local
    steps: NATURAL
  do
    if (is_normal_mode and (handle_status = is_handle_up)) then
      from
        steps := 0
      until
        (not (is_normal_mode and (handle_status = is_handle_up))) or
        (gear_status /= is_gear_extending) or
        (steps = 1)
      loop
        main
        steps := steps + 1
      end
      check
        (not (is_normal_mode and (handle_status = is_handle_up))) or
        (gear_status /= is_gear_extending)
      end
    end
  end
\end{lstlisting}
\caption{Translation of the ``r21" requirement.}
\label{fig:r21_translation}
\end{figure*}

\begin{figure*}
\begin{lstlisting}
  r11_rs
  -- ((gear_status = is_gear_extended) and (door_status = is_door_closed)) keeps holding under
  -- (is_normal_mode and (handle_status = is_handle_down))
    do
      if ((is_normal_mode and (handle_status = is_handle_down)) and
         ((gear_status = is_gear_extended) and (door_status = is_door_closed))) then
        main
        check
          ((is_normal_mode and (handle_status = is_handle_down)) implies
            ((gear_status = is_gear_extended) and (door_status = is_door_closed)))
        end
      end
    end
\end{lstlisting}
\caption{Translation of the ``r11\_rs" requirement.}
\label{fig:r11_bis_response_stability_translation}
\end{figure*}

Translation of requirement \e{r11_bis} (\autoref{fig:r11_bis_translation}) is an application of the \e{response_holds_within_k_steps} pattern (\autoref{fig:response_maximal_distance}), where:
\begin{itemize}
\item \e{stimulus} equates to:
\begin{lstlisting}
is_normal_mode and
  (handle_status=is_handle_down)
\end{lstlisting}

\item \e{response} equates to:
\begin{lstlisting}
(not (is_normal_mode and
  (handle_status=is_handle_down))) or
((gear_status=is_gear_extended) and
  (door_status=is_door_closed))
\end{lstlisting}

\end{itemize}

The idea behind the response is that there may be two reasons for the gear not to extend and the door not to close:
\begin{itemize}
\item An abnormal situation that leads to quitting the normal mode.
\item The crew changes their mind and pushes the handle up.
\end{itemize}


\section{Related work}

Modeling of real-time computation and related requirements is a well-investigated matter \cite{Yamada62}. Representation of real-time requirements, expressed in general or specific form, is a challenging task that has been attacked by the use of several formalisms both in sequential and concurrent settings, and in a broad set of application domains. The difficulty (or impossibility) to fully represents general real-time requirements other than in natural language, or making use of excessively complicated formalisms (unsuitable for software developers), has been recognized. 

In \cite{Mazzara10} the domain of real-time reconfiguration of system is discussed, emphasizing the necessity of adequate formalisms. The problem of modeling real time in the context of services orchestration in Business Process, and in presence of abnormal behavior has been examined in \cite{Mazzara05} and \cite{FerrucciBM14} by means, respectively, of process algebra and temporal logic. Modeling of protocols also requires real-time aspects to be represented \cite{BergerH00}. Event-B has also been used as a vector for real-time extension \cite{Iliasov2012} in order to handle embedded systems requirements. 

In all these studies, the necessity emerged of focusing on specific typology of requirements using ad-hoc formalisms and techniques, and making use of abstractions. The notion of ``real-time" is often abstracted as \textit{number of steps}, a metric commonly used. In this paper we follow the same approach, inheriting both strength (simplicity of the model and effectiveness for applicative purposes) and limitations (temporal logic and time automata themselves miss to capture a precise notion of \textit{real-time}).

\section{Conclusions and future work}
\label{sec:conclusions}


Software developers reason in an \textit{imperative/operational} manner. This claim is supported both by anecdotal experience and by empirical evidence \cite{Fahland2009}. 
Requirements expressed in imperative/operational fashion would therefore results of easier comprehensions for developers and would simplify the process of negotiation behind requirements elicitation. 

In the method described in this paper, requirements are expressed in a formalism (or language) that seamlessly stay the same along the whole process, without the need of switching between different instruments or mental paradigms. At the same time, the linguistic tool used to define them also allows for automatic verification of correctness.

The meaning of correctness here remains subject to the assumption that requirements engineers and stakeholders agree on a list of desiderata that is indeed the intended one. Assuming a non-faulty process of intention transferring (and this assumption is common to any other approach too), requirements are now more easily manageable by software engineerings all the way from elicitation to verification.  

The result of elicitation process is a set of requirements in natural language. The full realization of the presented method would imply an automatic (or semi-automatic) translation from natural language into a structured representation that, although completely intuitive for software developers, it is possibly not easy to manage for average stakeholders. The first part of this process, i.e., the translation from natural language into the current representation (and back) is under development. A tool automatically translates semi-structured natural language into the Hoare-triple-based representation \cite{Bormotova17}, allowing also the opposite direction, i.e. back to natural language \cite{Skukov17}, so that software engineers would be able to negotiate back requirements with stakeholders using a format they would comprehend. The role of the requirement engineers would then consist in concluding the elicitation phase with a set of requirements in semi-structured natural language, which the tool would be able to process in an entirely automatic manner.

This paper supports the idea of seamless development describing a method supported by a formalism that stay the same along the whole process, from requirements to deployment. Alternative approaches have also been experimented which make use of formalism-based toolkits, where ad hoc notations are adopted for each development phase \cite{GmehlichGLIJM13}.

{{{
	\bibliographystyle{ieeetr}
	\bibliography{EncodingLTL}
}}}
\end{document}